\documentclass{moriond}

\def\Journal#1#2#3#4{{#1} {\bf #2}, #3 (#4)}

\def\PLB{{\em Phys. Lett.}  B}

\def\JHEP{{\em JHEP}}
\def\EPJC{{\em EPJC}}

\def\be{\begin{equation}}
\def\ee{\end{equation}}
\def\bea{\begin{eqnarray}}
\def\eea{\end{eqnarray}}

\newcommand{\Photo}{}

\begin{document}
\vspace*{4cm}
\title{$\mathrm{t\bar{t}+X}$ production at ATLAS and CMS}

\author{Nicolas Tonon on behalf of the ATLAS and CMS Collaborations}

\address{Deutsches Elektronen-Synchrotron (DESY) \\
Notkestra{\ss}e 85, 22607 Hamburg}

\maketitle\abstracts{Recent inclusive and differential cross section measurements of the associated production of top quark pairs with gauge bosons or heavy-flavor jets are reported. 
A search for physics beyond the standard model in the top quark sector is also presented.
All measurements are based on data samples of proton-proton collisions at $\sqrt{s}=13$~TeV collected by the ATLAS and CMS experiments at the CERN LHC. 
No significant deviation from the standard model predictions is observed.}

\section{Introduction}
The standard model (SM) of particle physics is supported by a large amount of experimental results covering a wide energy range up to the TeV scale accessible at the CERN LHC. 
However, it does not provide explanations for several key observations such as the existence of dark matter and dark energy, or the masses of neutrinos. More generally, there exist a number of indications that the SM only corresponds to a low-energy approximation to a more fundamental theory.

The large top quark mass of about 173 GeV corresponds to a Yukawa coupling to the Higgs boson close to unity. 
This suggests that the top quark may play a special role within the SM, and that its closer study may shed light on the electroweak symmetry breaking mechanism.
Most canonical top quark processes have now entered a regime of precision differential measurements at the LHC, and their uncertainties are systematics-dominated. 

Many theories beyond the SM predict sizable deviations in the electroweak couplings of the top quark with respect to SM predictions. These couplings can be probed at tree level using the associated production of top quark pairs with vector bosons. 
In addition, the study of the associated production of top quark pairs with heavy-flavor jets probes several QCD predictions and provides crucial insights on the modeling of the $\mathrm{t\bar{t}}$ system, and can be used to compare different MC generators.
It also constitutes a major background to analyses such as $\mathrm{t\bar{t}H \to b\bar{b}}$, which motivates further the precise characterization of these processes.

\section{$\mathrm{t\bar{t}Z}$ cross section measurements at ATLAS and CMS}
Inclusive and differential measurements of the $\mathrm{t\bar{t}Z}$ cross section were performed by ATLAS~\cite{ttz_atlas} and CMS~\cite{ttz_cms} in $3\ell$ ($\ell = \textrm{e,}\mu$) and $4\ell$ final states using data samples with integrated luminosities of $139~\mathrm{fb}^{-1}$ and $77.5~\mathrm{fb}^{-1}$, respectively. 

Both analyses fit the data simultaneously in multiple event categories that are defined based on the multiplicities and properties of leptons and jets in the events. Control regions enriched in WZ and ZZ events are included, and the ATLAS analysis treats the normalizations of these processes as free parameters. The inclusive CMS result $\sigma(\mathrm{t\bar{t}Z})=0.95 \pm 0.05~\mathrm{(stat.) \pm 0.06~\mathrm{(syst.)~pb}}$ and ATLAS result $\sigma(\mathrm{t\bar{t}Z})=1.05 \pm 0.05~\mathrm{(stat.)} \pm 0.09~\mathrm{(syst.)~pb}$ agree with each other, and reach a better precision ($8-10\%$) than that of the best available SM prediction at NLO+NNLL accuracy~\cite{pred_ttz}.

\begin{figure}[!hbtp]
\begin{minipage}{0.45\linewidth}
\centerline{\includegraphics[width=1.\linewidth]{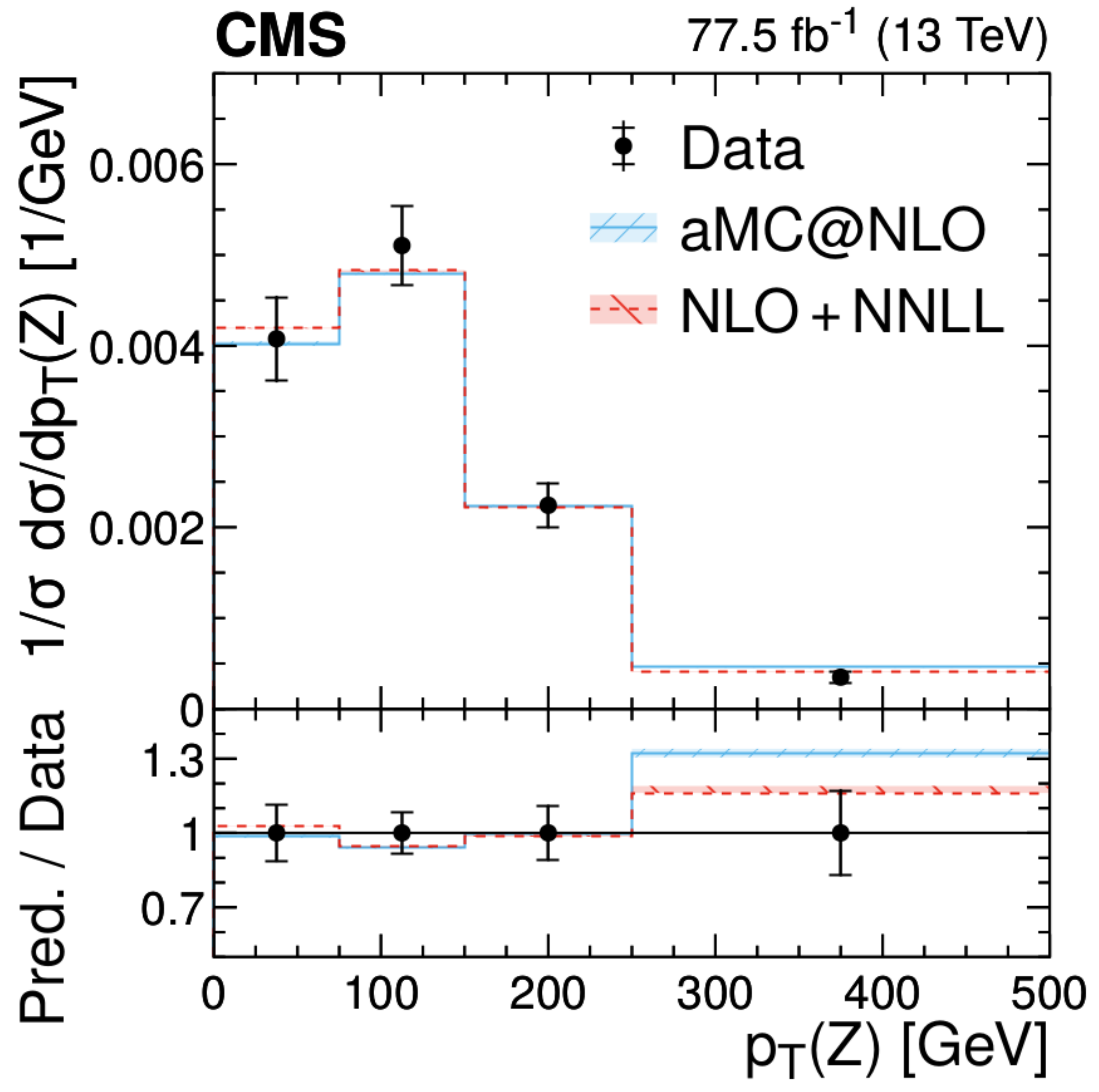}}
\end{minipage}
\hfill
\begin{minipage}{0.47\linewidth}
\centerline{\includegraphics[width=1.\linewidth]{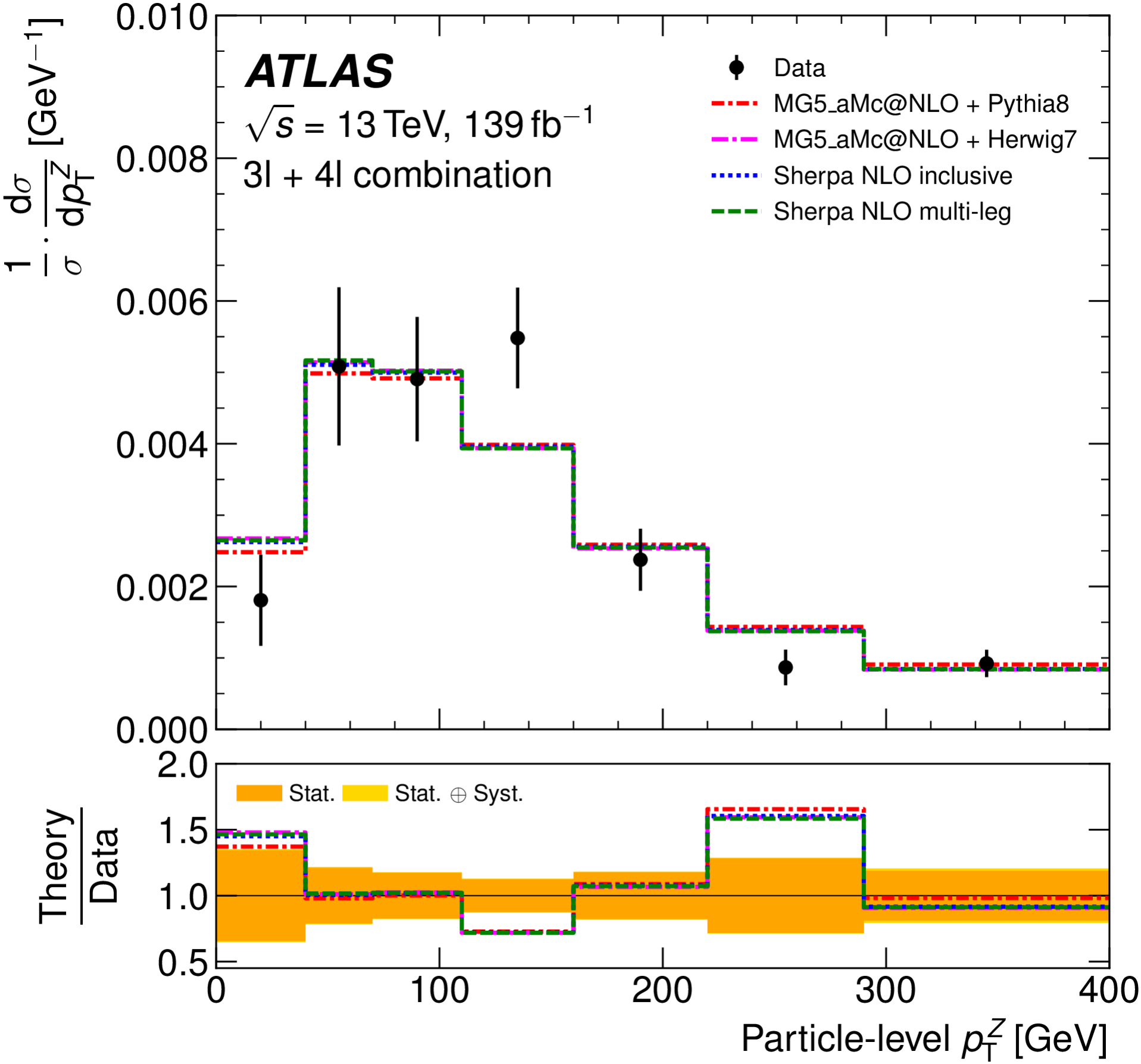}}
\end{minipage}
\caption[]{Normalized $\mathrm{t\bar{t}Z}$ cross section measurements as a function of $\mathrm{p_{T}(Z)}$ by CMS~\cite{ttz_cms} (left) and ATLAS~\cite{ttz_atlas} (right).}
\label{fig:ttz}
\end{figure}

The ATLAS analysis performs differential cross section measurements at the parton- and particle-level as a function of nine observables related to the Z boson, the $\mathrm{t\bar{t}}$ system, and the jet multiplicity. The CMS analysis provides parton-level differential measurements as a function of $\mathrm{p_{T}(Z)}$, and of the cosine of the angle between the direction of the Z boson in the detector reference frame and that of the negatively-charged lepton from the Z boson decay in the Z boson rest frame.
These two observables are also used to constrain anomalous $tZ$ couplings and Wilson coefficients (WCs) describing the interaction strengths of four relevant effective field theory (EFT) operators.
The normalized differential cross section measurements of $\mathrm{p_{T}(Z)}$ from both analyses are shown in Fig.~\ref{fig:ttz}. Good overall data-to-MC agreement is observed in differential distributions.

\section{$\mathrm{t\bar{t}\gamma}$ cross section measurements at ATLAS and CMS}
\begin{figure}[!hbtp]
\begin{minipage}{0.45\linewidth}
\centerline{\includegraphics[width=1.\linewidth]{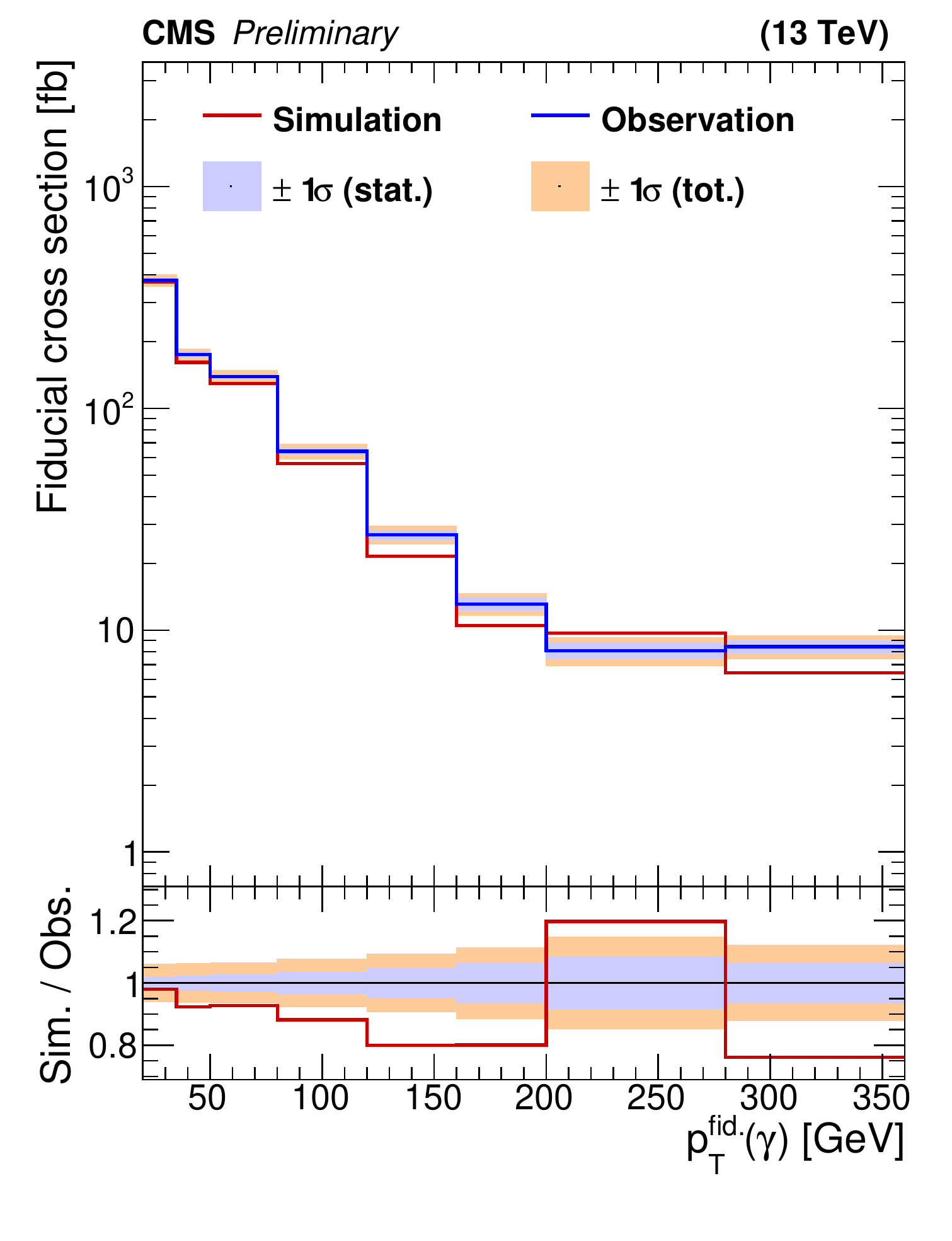}}
\end{minipage}
\hfill
\begin{minipage}{0.57\linewidth}
\centerline{\includegraphics[width=1.\linewidth]{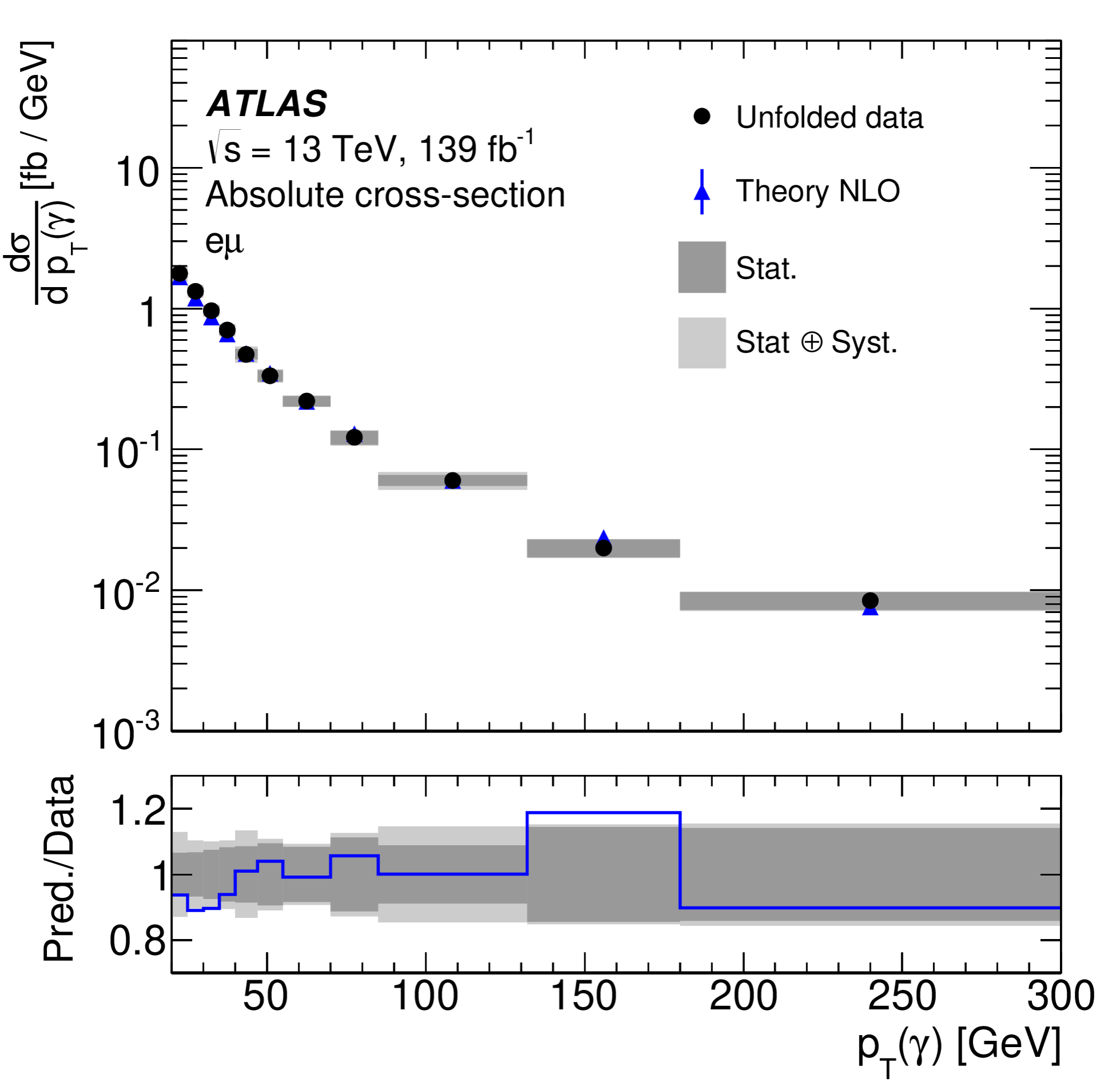}}
\end{minipage}
\caption[]{Absolute $\mathrm{t\bar{t}\gamma}$ cross section measurements as a function of $\mathrm{p_{T}(\gamma)}$ by CMS~\cite{ttg_cms} (left) and ATLAS~\cite{ttg_atlas} (right).}
\label{fig:ttg}
\end{figure}

Inclusive and differential measurements of the $\mathrm{t\bar{t}\gamma}$ cross section were performed by ATLAS~\cite{ttg_atlas} and CMS~\cite{ttg_cms} using the full Run~2 dataset. 
The ATLAS analysis targets the high-purity $\mathrm{e}\mu$ channel and considers both the $\mathrm{t\bar{t}\gamma}$ and $\mathrm{tW\gamma}$ signals. The simulation of the signal samples includes resonant, non-resonant, interference, and off-shell effects, which allows for a direct comparison with NLO predictions for the $\mathrm{pp \to bWbW\gamma}$ process~\cite{pred_ttg1,pred_ttg2}. A fit to the data is performed using the distribution of the scalar $\mathrm{p_{T}}$ sum of all leptons, photons, jets, and missing transverse momentum in the events.
The CMS analysis targets the $\ell\mathrm{+jets}$ channel, constrains the major backgrounds in-situ using dedicated sidebands, and performs a simultaneous fit in 46 event categories. Both inclusive results are consistent with SM predictions, and reach a comparable precision of about $6\%$. Both analyses provide differential measurements as a function of several observables, including $\mathrm{p_{T}(\gamma)}$ as shown in Fig.~\ref{fig:ttg}, and good overall data-to-MC agreement is observed. In addition, the CMS analysis sets constraints on two EFT operators impacting the $t-\gamma$ coupling.

\section{Search for physics beyond the standard model using the associated top quark production at CMS}
A CMS analysis~\cite{top19001} using a novel approach to parameterize EFT effects at the detector-level constrained simultaneously 16 WCs using $41.5~\mathrm{fb}^{-1}$ of data. 
It considered EFT effects in five associated production modes of the top quark with gauge and Higgs bosons ($\mathrm{t\bar{t}Z}$, $\mathrm{t\bar{t}W}$, $\mathrm{tZq}$, $\mathrm{t\bar{t}H}$, $\mathrm{tHq}$) in multilepton final states, and performed counting experiments in event categories enriched in different processes. 
Two standard deviation confidence intervals were computed for each WC while either fixing other WCs to zero or profiling them.
This represents an important step towards direct measurements including EFT effects simultaneously in all relevant processes.

\section{Associated $\mathrm{t\bar{t}}$ production with heavy-flavor jets at ATLAS and CMS}
The $\mathrm{t\bar{t}b\bar{b}}$ cross section was measured using $36~\mathrm{fb}^{-1}$ of data by ATLAS in the $\mathrm{e}\mu$ and $\ell\mathrm{+jets}$ channels, and by CMS in the $2\ell$, $\ell\mathrm{+jets}$, and all-hadronic channels.
All these analyses extract results by performing a fit to the b-tagging discriminants of the jets identified as originating from the additional $\mathrm{b\bar{b}}$ pair.

\begin{figure}[!hbtp]
\begin{minipage}{0.50\linewidth}
\centerline{\includegraphics[width=1.15\linewidth]{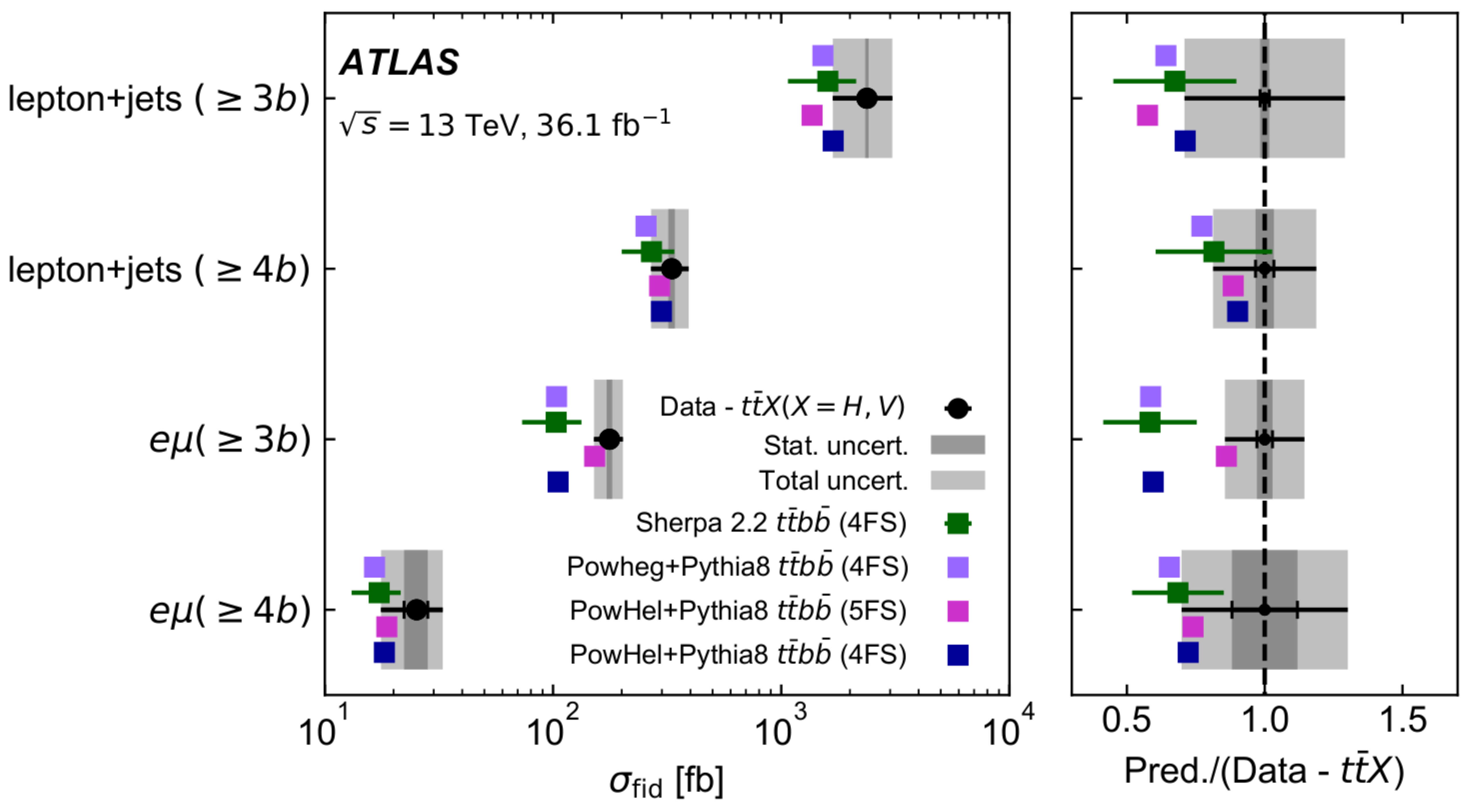}}
\end{minipage}
\hfill
\begin{minipage}{0.50\linewidth}
\centerline{\includegraphics[width=0.85\linewidth]{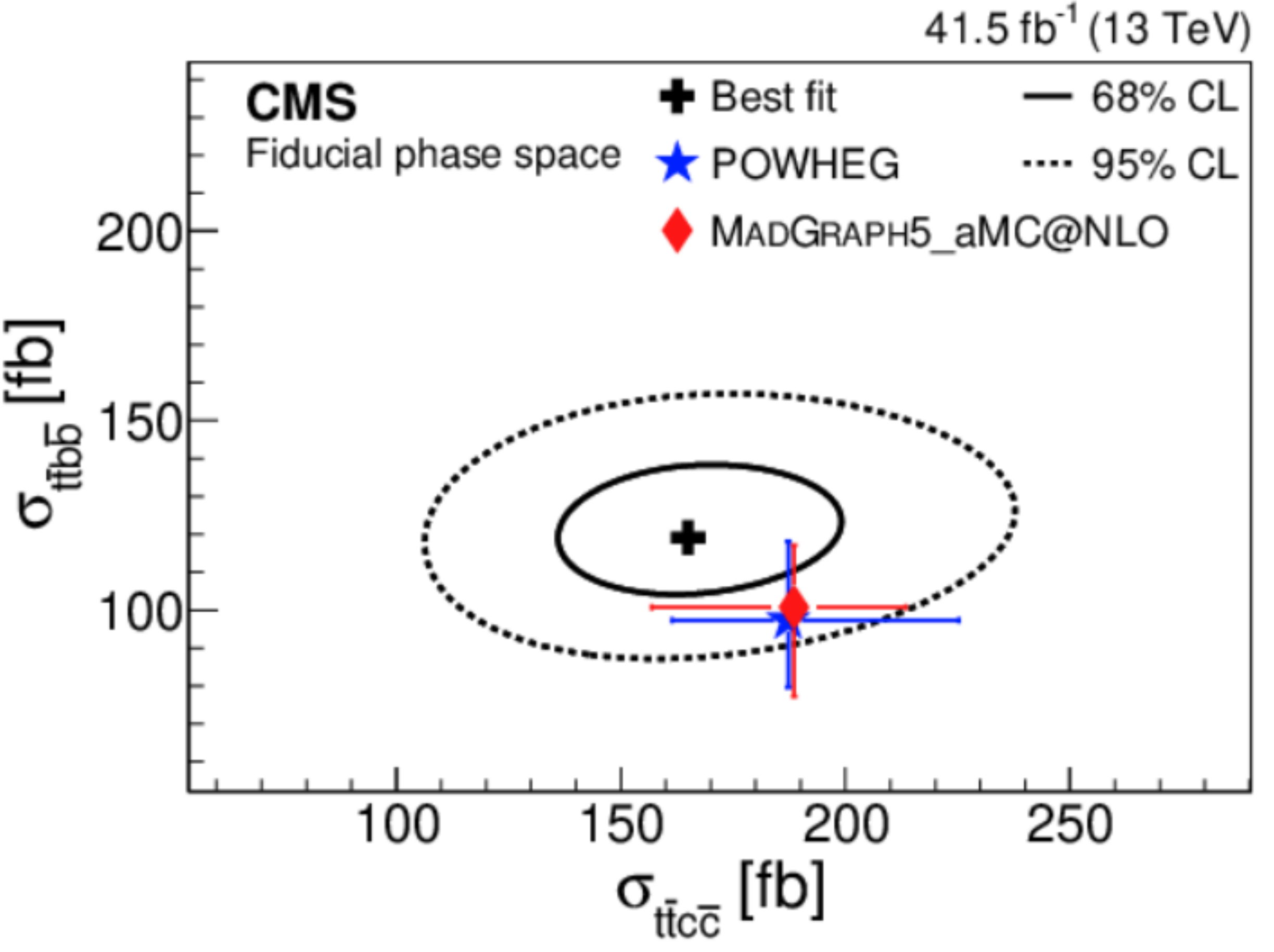}}
\end{minipage}
\caption[]{Left: comparison of fiducial $\sigma(\mathrm{t\bar{t}b\bar{b}})$ measurements by ATLAS with predictions from several generators~\cite{tthf_lep_atlas}. Right: result of a two-dimensional scan of $\sigma(\mathrm{t\bar{t}c\bar{c}})$ and $\sigma(\mathrm{t\bar{t}b\bar{b}})$ by CMS~\cite{ttcc_cms}.}
\label{fig:tthf}
\end{figure}

In the ATLAS analysis~\cite{tthf_lep_atlas}, the contributions from the non-$\mathrm{t\bar{t}}$ and $\mathrm{t\bar{t}H/Z} \to \mathrm{t\bar{t}b\bar{b}}$ backgrounds are substracted based on MC simulations. Cross sections are computed in different fiducial phase spaces as illustrated in Fig.~\ref{fig:tthf} (left). Relative precisions of $17\%$ and $13\%$ are achieved in the $\ell\mathrm{+jets}$ and $\mathrm{e}\mu$ channel, respectively, and several differential measurements are unfolded at particle-level.
The CMS leptonic analysis~\cite{tthf_lep_cms} measures $\sigma(\mathrm{t\bar{t}b\bar{b}})$, $\sigma(\mathrm{t\bar{t}j\bar{j}})$, and their ratio. Fiducial cross sections extracted in the $\ell\mathrm{+jets}$ and $2\ell$ channels reach precisions of $12\%$ and $13\%$, respectively.
The CMS all-hadronic analysis~\cite{tthf_jets_cms} benefits from a large branching ratio and a fully-reconstructible final state, but suffers from huge QCD backgrounds. The contribution from the latter is strongly reduced by using multivariate analysis techniques (quark-gluon likelihood discriminant, jet-assignment BDT, and weakly-supervised BDT to separate $\mathrm{t\bar{t}}$ and QCD processes).
Although no significant deviation is observed with respect to SM predictions, all three analyses consistently report a trend of under-prediction of $\sigma(\mathrm{t\bar{t}b\bar{b}})$ by all MC generators, which motivates further differential studies based on the full Run~2 data sample.

Finally, CMS reported the first-ever measurement of the $\mathrm{t\bar{t}c\bar{c}}$ cross section~\cite{ttcc_cms}, which is particularly challenging due to the difficulty to identify c-jets. This analysis exploits a novel c-tagger algorithm calibrated in a dedicated control region~\cite{ttcc_calib}. Along with the response of a neural network (NN) trained to correctly assign jets in the events, c-tagging jet discriminants are provided as input to a multiclass NN classifier tasked with separating the $\mathrm{t\bar{t}c\bar{c}}$ and $\mathrm{t\bar{t}b\bar{b}}$ processes. A fit to the response of this NN is performed to measure $\sigma(\mathrm{t\bar{t}c\bar{c}})$, $\sigma(\mathrm{t\bar{t}b\bar{b}})$, $\sigma(\mathrm{t\bar{t} + light~jets})$, as well as their ratios to the inclusive $\sigma(\mathrm{t\bar{t}jj})$ value. 
All results are consistent with POWHEG predictions within two standard deviations. The measured value for $\sigma(\mathrm{t\bar{t}c\bar{c}})$ reaches a precision of $20\%$ and is slightly over-predicted, while that for $\sigma(\mathrm{t\bar{t}b\bar{b}})$ slightly exceeds the prediction as shown in Fig.~\ref{fig:tthf} (right).

\section{Summary}
Recent inclusive and differential cross section measurements of the associated production of top quark pairs with gauge bosons or heavy-flavor jets were reported. 
While no significant deviation from the standard model predictions is observed, three analyses based on data collected in 2016 by ATLAS and CMS consistently report a trend of under-prediction of $\sigma(\mathrm{t\bar{t}b\bar{b}})$ by all MC generators.
A search for physics beyond the standard model targeting five associated production modes of the top quark was also presented.


\end{document}